# Interface Dzyaloshinskii-Moriya interaction in the interlayer exchange antiferromagnetic coupled Pt/CoFeB/Ru/CoFeB systems

M. Belmeguenai [1,*], H. Bouloussa[1], Y. Roussigné[1], M. S. Gabor[2], T. Petrisor jr.[2], C. Tiusan[2,3], H. Yang[4], A. Stashkevich[1], and S. M. Chérif[1].

[1]*LSPM, CNRS-Université Paris 13, 99 avenue Jean-Baptiste Clément Université Paris 13, 93430 Villetaneuse, France*
[2]*Center for Superconductivity, Spintronics and Surface Science, Physics and Chemistry Department, Technical University of Cluj-Napoca, Str. Memorandumului No. 28 RO-400114 Cluj-Napoca, Romania*
[3]*Institut Jean Lamour, CNRS, Université de Lorraine, 54506 Vandoeuvre, France*
[4]*Department of electrical and computer engineering, National University of Singapore, 117576, Singapore*

***Abstract-*** Interfacial Dzyaloshinskii-Moriya interaction (iDMI) in interlayer exchange coupled (IEC) Pt/$Co_{20}Fe_{60}B_{20}$(1.12 nm)/Ru/$Co_{20}Fe_{60}B_{20}$(1.12 nm) systems have been studied theoretically and experimentally. Vibrating sample magnetometer has been used to measure their magnetization at saturation and their interlayer exchange coupling constants. These latter are found to be of an antiferromagnetic nature for the investigated Ru range thickness (0.5-1 nm). Their dynamic magnetic properties were studied using Brillouin light scattering (BLS) technique. The BLS measurements reveal pronounced non-reciprocal spin waves propagation. In contrast to the calculations for symmetrical IEC CoFeB layers, this experimental non-reciprocity is Ru thickness and thus coupling strength dependent. Therefore, to explain the experimental behaviour, a theoretical model based on the perpendicular interface anisotropy difference between the bottom and top CoFeB layers has been developed. We show that the Ru thickness dependence of the spin wave non-reciprocity is well reproduced by considering a constant iDMI and different perpendicular interfacial anisotropy fields between the top and bottom CoFeB layers. This anisotropy difference has been confirmed by the investigation of the CoFeB thickness dependence of effective magnetization of Pt/CoFeB/Ru and Ru/CoFeB/MgO individual layers, where a linear behaviour has been observed.

\* belmeguenai.mohamed@univ-paris13.fr,

# I- Introduction

The exchange interaction plays an important role in magnetism and therefore is responsible of several phenomena in magnetic materials. This interaction can be direct (involving an overlap of electron wave functions from the neighboring atoms and Coulomb electrostatic interaction) or indirect (little or no direct overlap between neighboring electrons and mediated through an intermediary atoms). The direct exchange interaction between electrons arises from the Coulomb interaction and is responsible for the microscopic magnetic behavior. It may contain symmetric and asymmetric terms. The symmetric term, commonly known as the Heisenberg [1] interaction, usually leads to collinear magnetic structures. The asymmetric exchange, referred to as the Dzyaloshinskii–Moriya interaction [2, 3] (DMI), favors canted neighboring spins leading to various magnetization structures at the nanoscale such as helices [4] and skyrmions [5-7]. It changes the static and dynamic properties of domain walls [8] and leads to different energy (non-reciprocity) of two spin waves (SW) having the same wavelength and propagating along two opposite directions [9]. It is manifested by a difference between the frequencies of these two SWs. The DMI constant determination is thus reduced to this simple frequency difference measurement. Several experimental methods [10-12], largely based on how this interaction alters the properties of domain walls, were employed recently but Brillouin light scattering (BLS) spectroscopy remains the most direct method for DMI characterization. This scheme is simple, efficient, reliable and straightforward since few parameters are required for the experimental data fit [13, 14]. It also allows the investigation of both in-plane and perpendicular spontaneously magnetized films in contrast to domain wall techniques. DMI can be induced by a lack of inversion symmetry of the compound and a strong spin-orbit coupling. This can be achieved by using heavy metal/ferromagnet (HM/FM) heterostructures, giving rise to interfacial DMI.

Indirect exchange interactions such as coupling between two magnetic layers separated by a non-magnetic spacer layer is mediated by conduction electrons of the spacer layer which are scattered successively by the magnetic layers. The coupling, which oscillates in sign as a function of the thickness of the spacer layer [15, 16] was first observed by Grünberg [17] for transition metal systems. It is crucial for many applications in modern magnetic storage devices and spin electronics [18]. In practice, an antiferromagnetic interlayer coupling is easily revealed and measured by performing a magnetization measurement as a function of an applied magnetic field. A ferromagnetic coupling is much more difficult to detect and measure quantitatively by these static techniques, since the application of an external magnetic field has no direct action on the mutual orientations of the magnetizations of the successive magnetic layers. Therefore, dynamic methods such as ferromagnetic resonance and Brillouin light scattering (BLS) remain the most powerful and used means for the precise characterization of both coupling types. Indeed, in these methods and in analogy with coupled harmonic oscillators, the magnon modes in two magnetic films coupled via a nonmagnetic interlayer can be classified into acoustic and optic modes depending on whether the two film magnetizations precess in phase or out of phase, respectively [19]. The behavior of the spin-wave frequencies as a function of applied fields provides a great deal of information about the magnitude and functional form of the coupling energy.

Recently, Chen *et al.* [20] demonstrates an experimental approach to stabilize a room temperature skyrmion ground state in chiral magnetic films via interlayer exchange coupling (IEC). Indeed, Shawn *et al.* [21] have reported on the direct imaging of chiral spin structures including skyrmions in an exchange coupled thick ferromagnetic Co/Pt multilayer at room temperature with Lorentz transmission electron microscopy. Moreover, it is of utmost importance to investigate the spin waves spectrum in the presence of both DMI and IEC. Therefore, both experimental and theoretical investigations of this aspect will be reported in this work. We thus use BLS combined with vibrating sample magnetometry (VSM) to measure the combined effects of the IEC strength and of the DMI constant on SW non-reciprocity in Pt/Co$_{20}$Fe$_{60}$B$_{20}$/Ru/ Co$_{20}$Fe$_{60}$B$_{20}$. We show that although the two ferromagnetic (FM) layers are similar with the same thickness, caution should be paid to the interpretation of the SWs non-reciprocity. Indeed, the frequency difference between the two counter propagating SWs, usually attributed to DMI is also IEC strength dependent when the two FM layers present different perpendicular surface anisotropies.

## II- Samples and experimental techniques

A series of Co$_{20}$Fe$_{60}$B$_{20}$(1.12 nm)/Ru($t_{Ru}$)/Co$_{20}$Fe$_{60}$B$_{20}$(1.12 nm) multilayers ($t_{Ru}$ = 0.5, 0.6, 0.8 and 1 nm) have been grown by a sputtering magnetron system at room temperature on a thermally oxidized Si substrate. Prior to the deposition of the multilayer, a Ta(3 nm)/Pt(3 nm) buffer bilayer was deposited on the substrates. Finally, the trilayer were coated by a bilayer of MgO(1 nm)/Ta(3 nm). In this system, the Pt bottom layer induces perpendicular magnetic anisotropy and DMI whereas the Ru spacer layer is thought to only induce perpendicular anisotropy and to ensure IEC. The Ru thickness had been chosen in order to induce antiferromagnetic IEC between CoFeB layers. In order to determine the interface perpendicular anisotropy and the DMI constants, the individual layers Ta(3 nm)/Pt(3nm)/Ru(0.8nm)/CoFeB($t_{CFB}$)/MgO(1 nm)/Ta(3 nm) and Ta(3 nm)/Pt(3 nm)/CoFeB($t_{CFB}$)/Ru(0.8 nm)/Ta(3 nm) of variable CoFeB thicknesses (0.9 nm≤$t_{CFB}$≤5 nm) have also been grown in the same conditions.

VSM has been used to measure the hysteresis loops of the samples with the field applied parallel to the sample plane and to obtain the intrinsic value of the magnetization at saturation ($M_s$). The BLS technique gives access to SW modes as well as phonons with nonzero wave-vector values. In the BLS set-up, the SW, of a wave number ($k_{sw}$) in the range 0–20 μm$^{-1}$ (depending on the incidence angle $\theta_{inc}$: $k_{sw} = \dfrac{4\pi}{\lambda}\sin(\theta_{inc})$ in backscattering configuration), are probed by illuminating the sample with a laser having a wavelength λ=532 nm. The spectrometer is a JR Sandercock product based on a tandem Fabry-Perot interferometer. In order to select the SW lines, a crossed polarizer is placed on the path of the back scattered light from the sample. The magnetic field is applied perpendicular to the incidence plane, which allows for probing SWs propagating along the in-plane direction perpendicular to the applied field *i.e.* the Damon-Eshbach (DE) geometry where the iDMI effect on the SWs propagation non-reciprocity is maximal [22]. For each angle of incidence, the spectra will be obtained after sufficiently counting photons to have well-defined spectra where the line position can be determined with accuracy better than 0.2 GHz. The Stokes (S, negative frequency shift relative to the incident light as a magnon was created) and anti-Stokes (AS, positive frequency shift relative to the incident light as a magnon was absorbed) frequencies, detected simultaneously will then be determined from Lorentzian fits to the BLS spectra. All the measurements presented here were carried out at room temperature.

**III- Results and discussions.**

VSM hysteresis loops for a CoFeB/Ru/CoFeB trilayer with different Ru layer thicknesses are shown in figure 1. These loops clearly show that for all the samples, the magnetizations of the two CoFeB layers are antiferromagnetically coupled. Indeed, in zero applied magnetic fields, the magnetizations of successive magnetic layers are antiparallel to each other, resulting in zero remnant magnetization due to the antiferromagnetic interaction. When an external magnetic field is applied, the Zeeman energy tends to align the magnetizations of both layers in the field direction, so that the magnetizations progressively increase until a saturation field is reached. This saturation field is Ru thickness dependent as shown in figure 1. The in-plane saturation field allows for deriving the antiferromagnetic coupling constants. Since the variation of the magnetization versus the applied field before saturation is linear (figure 1), only the bilinear coupling constant $J_1$ has to be considered. Assuming in-plane magnetizations, one writes the energy per unit area as:

$$E = -t_{CFB} H M_s \cos(\varphi_{M1} - \varphi_H) - t_{CFB} H M_s \cos(\varphi_{M2} - \varphi_H) - J_1 \cos(\varphi_{M1} - \varphi_{M1}) \tag{1}$$

In the above expression, the in-plane anisotropy has been neglected, $\varphi_{M1}$, $\varphi_{M2}$ and $\varphi_H$, respectively, represent the in-plane (referring to the substrate edge) angles defining the direction of the magnetization of the two CoFeB layers and of the applied magnetic field. The energy E is minimal for $\varphi_{M1} = \varphi_{M2} = \varphi_H$ if $H > -2J_1/(t_{CFB}M_s)$ ; the in-plane saturation field is thus $H_{sat} = -2J_1/(t_{CFB}M_s)$. Using the $M_s$ value ($M_s$ = 1200 emu/cm$^3$, measured by VSM) and the saturation fields deduced from hysteresis loops shown in figure 1, the corresponding coupling constants are $J_1$=-0.45, -0.2, -0.14 and -0.0013 erg/cm$^2$, respectively for $t_{Ru}$= 0.5, 0.6, 0.8 and 1 nm. Note the very weak value of the antiferromagnetic interaction for 1 nm thick Ru spacer.

The typical BLS spectra are displayed in figure 2 for two Ru thicknesses at $k_{sw}$ = 20.45 and 8.08 µm$^{-1}$ and for two in-plane applied fields sufficient to saturate the magnetizations. Two main features are noticeable: one line (mode 1) is observable in the S and AS parts of each spectrum; the positions of these lines are not symmetrical. As the structure is made of two coupled FM layers, one expects two magnetic modes (optic and acoustic modes) in Stokes and anti-Stokes parts, as mentioned above. Nevertheless, intensity calculations (similar to those in [23] and including iDMI boundary conditions of [24]) using the magnetic parameters deduced from the field dependence of the frequency modes (not shown here), the gyromagnetic factor of 30.13 GHz/T, measured by ferromagnetic resonance and iDMI effective constant $D_{eff}$ = -0.84 mJ/m² (deduced from BLS measurements of Pt/CoFeB/Ru which will be presented below) reveal that the intensity of the second line (optic mode) is very weak and thus experimentally unobservable. Indeed, in the case of similar coupled ferromagnetic films, the optic mode profile presents opposite signs in the ferromagnetic layers thus leading to a vanishing resultant. This is because the CoFeB layers have the same thickness and not very different perpendicular anisotropy fields (400 Oe), as shown on figure 3a for $t_{Ru}$=0.8 nm. To observe this second mode, the perpendicular anisotropy difference should be significant as shown in figure 3b, where spectra calculations are presented for $t_{Ru}$=0.8 nm but assuming an anisotropy field difference of 4 kOe.

The observed frequency difference between S and AS lines ($\Delta F = F_S - F_{AS}$, where $F_S$ and $F_{AS}$ are the frequencies of S and AS lines, respectively) of the acoustic mode (mode 1) should be related *inter alia* to iDMI. Nevertheless, simulations (close to the ones presented in [25] and complemented with iDMI boundary conditions of ref [24]) shown in figure 4a reveal that the frequency shift for symmetrical CoFeB layers with similar magnetic properties should be independent of the IEC strength and thus of the Ru thickness. It is noticeable that $\Delta F$ for mode 2 (not observed) is slightly IEC dependent but its variation would not be experimentally detected, according to figure 4a insert. Moreover, the $\Delta F$ value for mode 1 is the half of that for a single layer (according to [13], $\Delta F = \frac{2\gamma}{\pi M_s} D_{eff} k_{SW}$ = 1.72 GHz) and it turns out that the coupled layers behave like a single layer with a double thickness. For systems having the same FM layers thickness and different perpendicular anisotropies, $\Delta F$ is IEC dependent (figure 4b). For antiferromagnetic coupled layers (considered in figure 4b), the SW non reciprocity of the mode 1 presents a maximum in the vicinity of $J_1 = 0$.

The experimental $k_{sw}$ dependences of $\Delta F$ for various Ru thicknesses as well as those of the Ta(3 nm)/Pt(3 nm)/CoFeB(1.12 nm)/Ru(0.8 nm)/Ta(3 nm) and Ta(3 nm)/Pt(3nm)/Ru(0.8nm)/CoFeB(1.12 nm)/MgO(1 nm)/Ta(3 nm) individual layers are shown in figure 5a. Note the negative sign of $\Delta F$, the variation of its slope with the Ru thickness and its small value compared to that of the single CoFeB layer [Ta(3 nm)/Pt(3 nm)/CoFeB(1.8 nm)/Ru(0.8 nm)/Ta(3 nm)]. The effective iDMI constants ($D_{eff}$) of the individual layers, deduced from the slope of $k_{sw}$ dependences of $\Delta F$ [13] using the above mentioned magnetization at saturation and the gyromagnetic ratio values, are found to be -0.84 mJ/m$^2$ and -0.3 mJ/m$^2$, respectively for Pt/CoFeB(1.12 nm)/Ru (0.8 nm) and Pt/Ru(0.8 nm)/CoFeB(1.12 nm)/MgO. This iDMI constant of Pt/CoFeB(1.12 nm)/Ru(0.8 nm) is in good agreement with that obtained by Tacchi [26] and Di [14]. The smaller iDMI value of Pt/Ru(0.8 nm)/CoFeB(1.12 nm)/MgO suggests that the thin Ru layer partially screens the interaction between Pt and CoFeB atoms and to completely cancel iDMI. This is in agreement with Tacchi *et al.* [26] observations that indicate that not only the interface Pt atoms are involved in iDMI but at least 1 nm thick Pt layer is concerned. The screening effect *via* Ir and Au spacer between Pt and ferromagnetic layer has been reported by Robinson *et al.* [27]. The iDMI cancellation would occur for a spacer thickness of about 1 nm. Furthermore, the experimental observed IEC dependence of $\Delta F$ is thus an indication that the bottom and the upper CoFeB layers have a different perpendicular anisotropy as shown by simulations (figure 4b). Therefore, the experimental data have been fitted using the same iDMI parameter $D_{eff}$ = -0.84 mJ/m$^2$ (of the single CoFeB layer: Pt/CoFeB/Ru), the above mentioned values of $J_1$ and different anisotropy fields for the bottom and top CoFeB layers. For this, the experimental $k_{sw}$ dependence of the S and AS frequencies of each sample are fitted, as illustrated for example in figure 5b for $t_{Ru}$=0.8 nm and $\Delta F$ is then calculated. The $t_{Ru}$ dependence of these anisotropy fields is shown in figure 6a, where higher anisotropy fields have been observed for the top CoFeB with thinner Ru layers. As the Ru thickness increases, the anisotropy difference decreases and changes sign for Ru thickness around 1 nm. This anisotropy field difference is due to the perpendicular interface anisotropy induced by the different buffer and capping layers used here (Pt, Ru and MgO). This interface anisotropy has been confirmed experimentally by investigating the thickness dependence of effective magnetization ($4\pi M_{eff}$=$4\pi M_s$-$H_\perp$) of the individual ferromagnetic layers Ta(3

nm)/Pt(3nm)/Ru(0.8nm)/CoFeB($t_{CFB}$)/MgO(1 nm)/Ta(3 nm) and Ta(3 nm)/Pt(3 nm)/CoFeB($t_{CFB}$)/Ru(0.8 nm)/Ta(3 nm) of variable CoFeB thicknesses, shown in figure 6b. The $M_{eff}$ values have been deduced from the fit of the experimental field dependence of the uniform precession mode frequency measured *via* ferromagnetic resonance. Using the above mentioned value of $M_s$, the interface anisotropy was found to be 1.02 erg/cm$^2$ and 0.68 erg/cm$^2$ for Pt/Ru(0.8 nm)/CoFeB/MgO and Pt/CoFeB/Ru(0.8 nm), respectively.

This feature of different anisotropy fields is very important to correctly evaluate the iDMI parameter and explain the frequency mismatch. In order to understand how this anisotropy field difference affects the spin waves non-reciprocity, the profile of the perpendicular to the plane component of the magnetization *versus* the stack thickness for Pt/CoFeB(1.12 nm)/Ru(0.8 nm)/CoFeB(1.12 nm)/MgO has been simulated. Figure 7 shows the obtained profile over the stack depth for two different bilinear IEC constants ($J_1$=0 and $J_1$=-0.14 erg/cm$^2$) and for CoFeB films having similar ($H_\perp$=9.5 kOe for both CoFeB layers) or different anisotropy fields ($H_\perp$=7.5 kOe and $H_\perp$=11.5 kOe for bottom and top CoFeB layers, respectively) subjected to iDMI ($D_{eff}$ =0 or -0.84 mJ/m$^2$) under 5 kOe in-plane applied magnetic field. Note that 0 corresponds to the beginning of the top CoFeB layer. The profiles have been calculated using the fluctuation dissipation theorem like the line intensities calculation presented in figure 3. The displayed curves correspond to the square of the magnitude of the thermo-activated dynamic magnetization component perpendicular to the films. In order to understand the influence of both iDMI and anisotropy asymmetry on the frequency mismatch, the profile of the Stokes mode will be compared to the anti-Stokes one. If these profiles are the same when mirrored with respect to the median plane of the CoFeB/Ru/CoFeB stack, then the corresponding frequencies are equal. Different mirrored profiles imply frequency mismatch. Therefore the profiles asymmetry extent measures the frequency difference. Indeed, in the perfectly symmetric case with $D_{eff}$ = 0 and $H_\perp$=9.5 kOe for both CoFeB layers, whatever the exchange coupling value, the calculated profile of AS mode is obtained by symmetry with respect to the median plane from the calculated profile for S mode (figures 7a and 7d for $J_1$=0 and -0.14 erg/cm$^2$, respectively). Consequently, $F_S$ and $F_{AS}$ are equal. In the presence of iDMI ($D_{eff}$ =-0.84 mJ/m$^2$), the S and AS profiles are not anymore symmetric with respect to the median plane but this asymmetry (figure 7b for $J_1$=0) does not depend on the exchange coupling $J_1$ (figure 7e for $J_1$=-0.14 er/cm$^2$). Therefore, $F_S$ and $F_{AS}$ are different but the frequency difference does not vary with the $J_1$ value as already mentioned on figure 4a (mode 1). Finally, in the case of asymmetric magnetic anisotropies ($H_\perp$=7.5 kOe and $H_\perp$=11.5 kOe for bottom and top CoFeB layers, respectively) with iDMI, the S and AS profiles are not symmetric with respect to the median plane and this asymmetry is more pronounced for $J_1$ = 0 (figure 7c) when compared to that for $J_1$=-0.14 erg/cm$^2$ (figure 7f). This feature can be related to the frequency difference variation with the exchange coupling as presented on figure 4b. Finally, the influence of both iDMI and anisotropy asymmetry on the frequency difference can be explained by the effect of iDMI and anisotropy asymmetry on the eigen mode profiles.

**Conclusion**

This paper presents an experimental and theoretical study of stacks made of two coupled ferromagnetic layers deposited on a heavy metal inducing interfacial Dzyaloshinskii Moryia interaction. The static measurements prove that the coupled layers have spontaneous opposite magnetizations and enable to derive the coupling between the two ferromagnetic layers. The spin wave observations by means of Brillouin light scattering reveal a Stokes and anti-Stokes frequency difference that could be related to the interfacial Dzyaloshinskii Moryia interaction. Nevertheless the simulations show that the frequency difference is also influenced by the coupling between the ferromagnetic layers when they possess different anisotropies. Magnetization profile calculations allowed for explaining this frequency mismatch by including anisotropy field difference between the top and bottom ferromagnetic layers of the stack. Therefore, the interfacial Dzyaloshinskii Moryia interaction parameter is correctly derived once the different anisotropies are evaluated.

**Acknowledgements**

This work has been supported by the Conseil regional d'Île-de-France through the DIM NanoK (BIDUL project) and by USPC-NUS via the project IMANSA N° 2016-R/USPC-NUS.

**Figure 1: Belmeguenai *et al.*

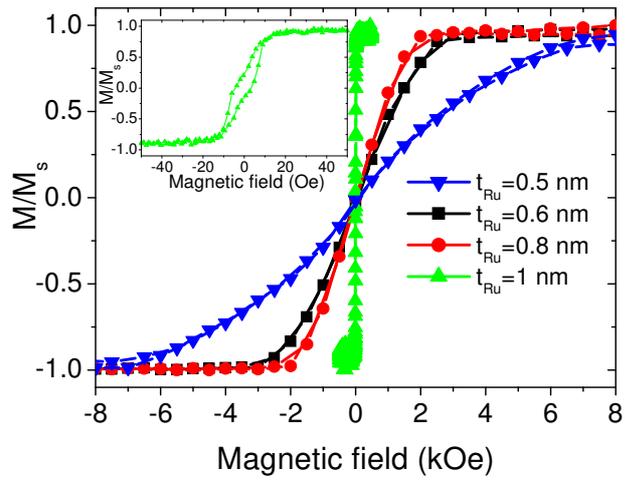

**Figure 2: Belmeguenai *et al.***

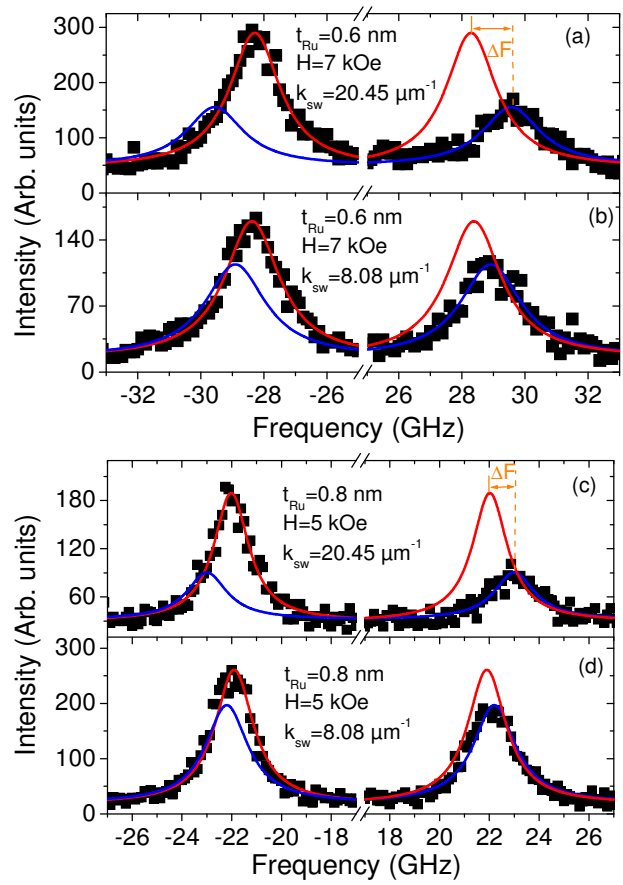

**Figure 3: Belmeguenai *et al.***

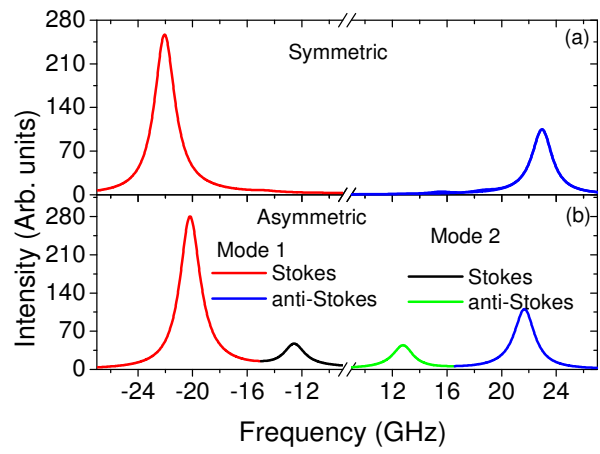

**Figure 4: Belmeguenai *et al.***

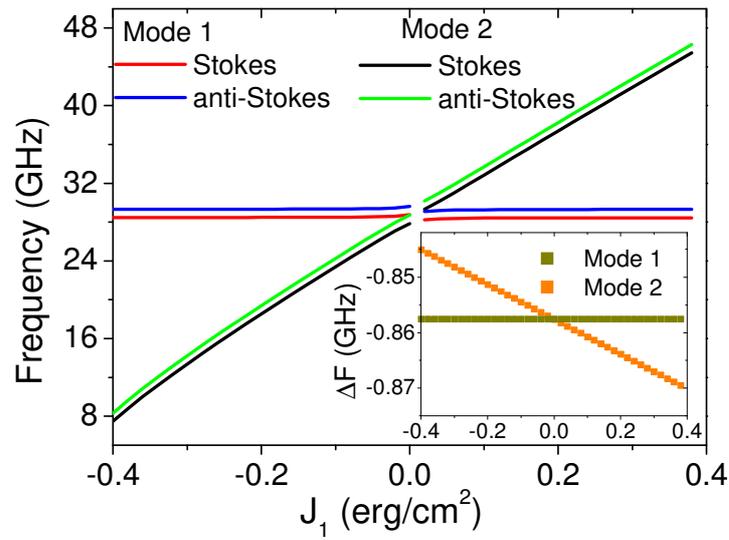

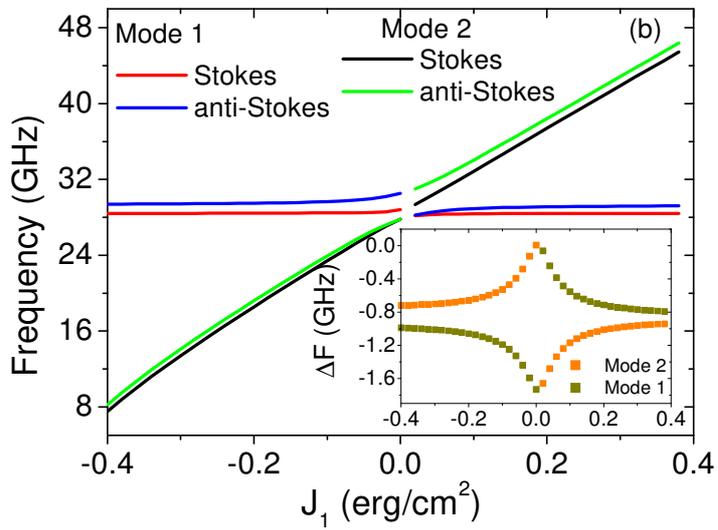

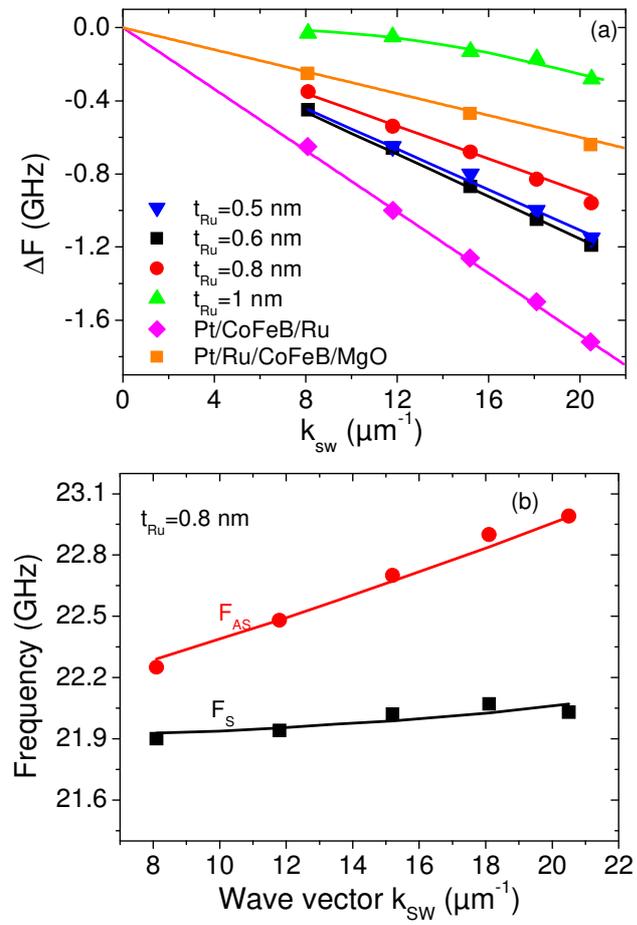

**Figure 6: Belmeguenai *et al.*

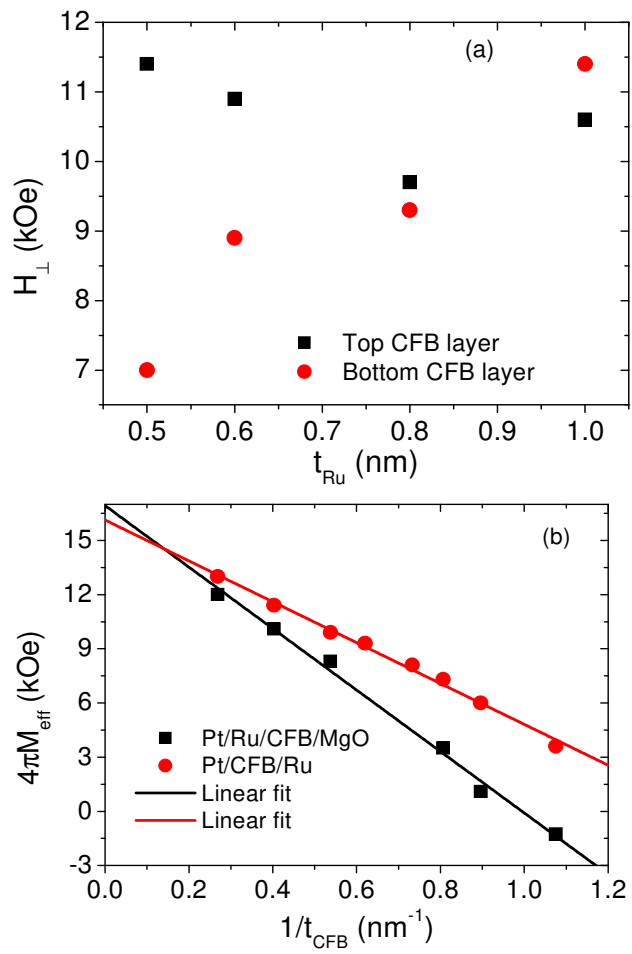

**Figure 7: Belmeguenai** *et al.*

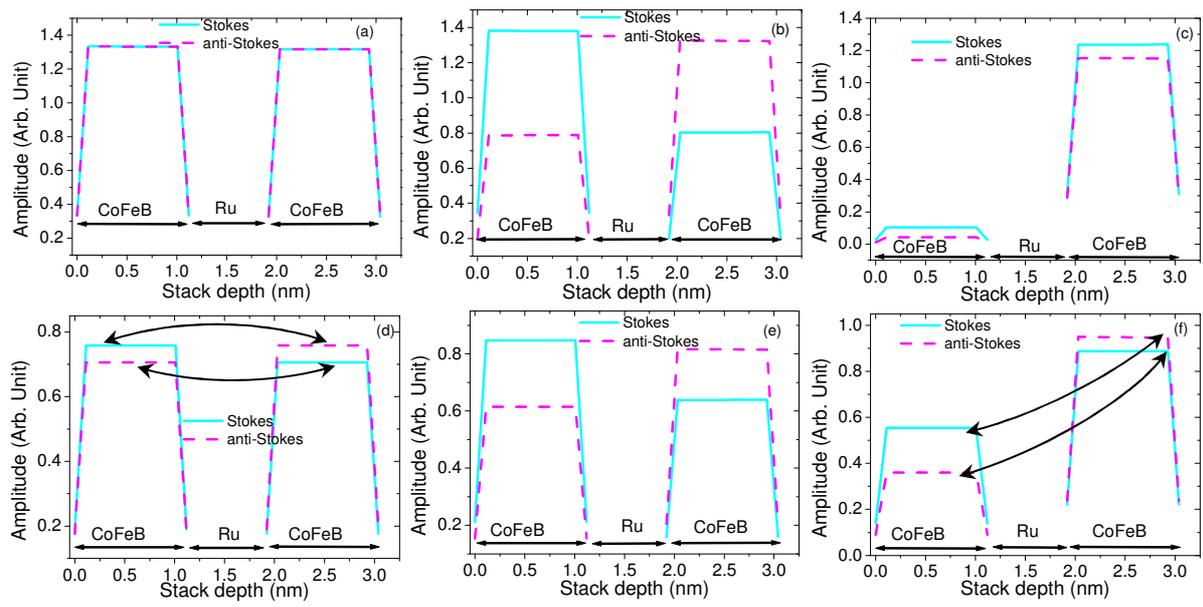

**Figure captions**

**Figure 1:** (Color online) VSM hysteresis loops showing the normalized in-plane magnetization component as function of the in-plane applied magnetic field for Pt/CoFeB(1.12 nm)/Ru ($t_{Ru}$)/CoFeB(1.12 nm)/MgO systems of various Ru thicknesses ($t_{Ru}$). The insert shows a zoom on the hysteresis loop for $t_{Ru}$=1 nm.

**Figure 2:** (Color online) BLS spectra measured for Pt/CoFeB(1.12 nm)/Ru ($t_{Ru}$)/CoFeB(1.12 nm)/MgO with two Ru thicknesses ($t_{Ru}$) at 7 kOe [(a) and (b)] and 5 kOe [(c) and (d)] in-plane applied magnetic field values and at two characteristic light incidence angles corresponding to $k_{sw}$= 8.08 and 20.45 $\mu m^{-1}$. Symbols refer to the experimental data and solid lines are the Lorentzian fits. Fits corresponding to negative applied fields are presented for clarity and direct comparison of the Stokes and anti-Stokes frequencies.

**Figure 3:** (Color online) Example of BLS spectra for Pt/CoFeB(1.12 nm)/Ru(0.8 nm)/CoFeB(1.12 nm)/MgO submitted to 5 kOe applied magnetic field. These simulations are obtained using our model presented in [20], including iDMI boundary conditions of [21], $M_s$ = 1200 emu/cm$^3$, $J_1$=-0.14 erg/cm$^2$, gyromagnetic factor of 30.13 GHz/T, $k_{sw}$=20.45 µm$^{-1}$, $D_{eff}$ = -0.84 mJ/m$^2$ and (a) $H_\perp$=9.3 kOe and $H_\perp$=9.7 kOe for bottom and top CoFeB layers, respectively. For (b) same parameters with $H_\perp$=9.3 kOe and $H_\perp$=13.3 kOe for bottom and top CoFeB layers, respectively have been used. Note the existence of two clearly intense modes in (b) due to the perpendicular anisotropy field difference between the top and the bottom CoFeB layers.

**Figure 4:** (Color online) Simulations showing the variation of the frequencies of the two excited modes (mode 1 and 2) in Pt/CoFeB(1.12 nm)/Ru/CoFeB(1.12 nm)/MgO *versus* the bilinear interlayer exchange coupling constant $J_1$ for in-plane magnetic applied 7 kOe, $k_{sw}$=20.45 µm$^{-1}$ and (a) $H_\perp$=9.4 kOe and $H_\perp$=9.4 kOe for bottom and top CoFeB layers, respectively. For (b) $H_\perp$=8.6 kOe and $H_\perp$=10.2 kOe for bottom and top CoFeB layers,

respectively. Simulations are based on the model in [22] complemented with iDMI boundary conditions of ref [21], using the other parameters of figure 3. Inserts of (a) and (b) are the frequency differences corresponding to modes 1 and 2.

**Figure 5:** (Color online) (a) Wave vector ($k_{sw}$) dependence of the experimental frequency difference $\Delta F$ of Pt/CoFeB(1.12 nm)/Ru ($t_{Ru}$)/CoFeB(1.12 nm)/MgO of various Ru thicknesses $t_{Ru}$ as well as those of the Pt/CoFeB(1.12 nm)/Ru(0.8 nm) and Pt/Ru(0.8 nm)/CoFeB(1.12 nm)/MgO. (b) $k_{sw}$ dependence of the experimental frequency of the observed mode (mode 1) of Pt/CoFeB(1.12 nm)/Ru (0.8 nm)/CoFeB(1.12 nm)/MgO for 5 kOe in-plane magnetic field and $k_{sw}$=20.45 µm$^{-1}$. Solid lines refer to fits using the model described in the paper, the anisotropy field differences shown in figure 6a and iDMI constants mentioned in the paper text.

**Figure 6:** (Color online) (a) Ru thickness dependence of the perpendicular anisotropy field deduced from fits of the experimental data of figure 5a. (b) CoFeB thickness dependence of the effective magnetization ($4\pi M_{eff}$) extracted from the fit of ferromagnetic resonance measurements of Pt/CoFeB($t_{CFB}$)/Ru(0.8 nm) and Pt/ Ru(0.8 nm)/CoFeB($t_{CFB}$)/MgO. Symbols refer to experimental data while solid lines are the linear fits.

**Figure 7:** (Color online) Profile of the perpendicular to the plane component of the thermo-activated dynamic magnetization *versus* the stack depth for Pt/CoFeB(1.12 nm)/Ru(0.8 nm)/CoFeB(1.12 nm)/MgO systems submitted to 5 kOe in-plane applied magnetic field. The first and second columns corresponds to calculations for CoFeB films having similar anisotropy fields ($H_\perp$=9.5 kOe for both CoFeB layers) while the third column refers to simulations for CoFeB films with different anisotropy fields ($H_\perp$=7.5 kOe and $H_\perp$=11.5 kOe for bottom and top CoFeB layers, respectively). Different cases corresponding to (a) $J_1$=0 and $D_{eff}$=0, (b) $J_1$=0 and $D_{eff}$=-0.84 mJ/m$^2$, (c) $J_1$=0 and $D_{eff}$=-0.84 mJ/m$^2$, (d) $J_1$=-0.14 erg/cm$^2$ and $D_{eff}$=0, (e) $J_1$=-0.14 erg/cm$^2$ and $D_{eff}$=-0.84 mJ/m$^2$ and (f) $J_1$=-0.14 erg/cm$^2$ and $D_{eff}$=-0.84

mJ/m$^2$ are considered. Note that for stack depth, 0 corresponds to the beginning of the top CoFeB layer. Arrows, indicated profiles which should be compared to understand the frequency mismatch.